\documentclass[a4paper,11pt]{article}
\usepackage{pos}
\usepackage{comment}

\title{H.E.S.S. follow-up of BBH merger events}
\ShortTitle{BBH mergers follow-up with H.E.S.S.}

\author*[a]{Halim Ashkar}
\author[a]{Francois Brun}
\author[b]{Clemens Hoischen}
\author[c]{Ruslan Konno}
\author[c]{Stefan Ohm}
\author[c]{Heike Prokoph}
\author[a]{Fabian Sch\"ussler}
\author[d]{Monica Seglar Arroyo}
\author[c]{Sylvia J Zhu}

\affiliation[a]{IRFU, CEA, Universit\'e Paris-Saclay, F-91191 Gif-sur-Yvette, France}

\affiliation[b]{Institut f\"ur Physik und Astronomie, Universit\"at Potsdam, Potsdam, Germany}

\affiliation[c]{DESY, D-15738 Zeuthen, Germany}

\affiliation[d]{Univ. Savoie Mont Blanc, CNRS,
Laboratoire d’Annecy de Physique des Particules - IN2P3,
74000 Annecy, France}

\forColl{H.E.S.S.} 

\emailAdd{contact.hess@hess-experiment.eu}

\abstract{We present here, follow-up observations of four Binary black hole BBH events performed with the High Energy Stereoscopic System (H.E.S.S.) in the Very High Energy (VHE) gamma-ray domain during the second and third LIGO/Virgo observation runs. Detailed analyses of the obtained data did not show significant VHE emission. We derive integral upper limit maps considering a generic $E^{-2}$ source spectrum in the most sensitive H.E.S.S energy interval ranging from 1 to 10 TeV. We also consider Extragalactic Background Light  absorption effects and derive integral upper limits over the full accessible energy range. We finally derive upper limits of the VHE luminosity for each event and compare them with the expected VHE emission from GRBs. These comparisons allow us to assess the H.E.S.S. gravitational wave follow-up strategies. For the fourth GW observing run O4, we do not expect to fundamentally alter our observing strategy, and will continue to prioritize sky coverage like for the previous runs.}

\FullConference{%
  *** 37th International Cosmic Ray Conference (ICRC2021), ***\\
  *** 12-23 July 2021 ***\\
  *** Berlin, Germany - Online ***
}

\begin{document}
\maketitle

\section{Introduction}
The detection of a short Gamma-ray Burst (GRB) emanating from the merger of two binary neutron stars (BNS), GW170817~\cite{GW170817, GW_GRB_170817}, provided definitive proof that such gravitational wave (GW) events can produce electromagnetic (EM) emission. The general consensus is that in order for EM emission to be produced in a compact binary coalescence (CBC) at least one of the objects should be a neutron star. 

In the case of binary black hole (BBH) mergers, the assumed lack of surrounding material makes it more difficult for EM emission to occur. However, coincidentally to the detection of the first GWs emanating from a BBH merger, GW150914~\cite{GW150914}, the Gamma-ray Burst Monitor instrument on board the \textit{Fermi} space observatory detected a weak gamma-ray transient~\cite{GW150914_grb}. This event sparked interest in the astrophysical community and triggered theoretical efforts for a model that could explain the observed transient. From these models, the existence of a circumbinary or remnant disk \cite{Perna2016_BBHEM, Murase2016_BBHEM, Kotera2016_BBHEM, Martin2018_BBHEM}, or of charged black holes \cite{Zhang2016_BBHEM, Fraschetti2018_BBHEM} are believed to be scenarios that could lead to potential EM emission. Furthermore, the possibility of EM emission from BBH happening in active galactic nuclei disks is also discussed~\cite{Graham2020_ZTFS190521g,Bartos2017_BBHEM_AGN}.

During the first three observation runs of Advanced LIGO and Advanced Virgo (O1, O2 and O3), 64 un-retracted GW event detection notices were sent. H.E.S.S., the High Energy Stereoscopic System, is an array of five Imaging Atmospheric Cherenkov Telescopes (IACTs) located in the Khomas Highland (Namibia). It is composed of four 12-m small telescopes and a 28-m large telescope, and is able to detect gamma rays in the VHE domain ranging from few tens of GeVs (depending on the zenith angle) to $\sim$ 100 TeV. During O2 and O3, H.E.S.S. observed six GW events; of these, one was the BNS merger GW170817, one was a neutron star — black hole merger,S200115j~\cite{S200115j}, and four were BBH mergers: GW170814~\cite{GW170814,O2_paper_catalog}, S190512at,  S190728q\cite{O3a_paper_catalog} and S200224ca\cite{S200224ca}. GW170817 was targeted during a short-term campaign discussed in~\cite{GW170817_HESS} and a long-term campaign discussed in~\cite{EM170817_HESS}. In both campaigns, no significant VHE signal was detected. Due to bad weather, S200115j was only observed with one observation with less than a 1\% chance of having observed the true location of the event.Therefore, this observation is not discussed here. 

In this contribution, we report on the search for VHE emission emanating from the four BBH merger events. The aim is to verify the existence or absence and to constrain VHE emission from BBH mergers. Moreover, the search and analysis methods presented here are used for all types of GW event follow-ups with H.E.S.S. Therefore, in order to prepare for the fourth observation run, O4, it is important to assess the sensitivity and the overall current H.E.S.S. observation strategy to look for potential improvements and changes. 


\section{Observations summary}
\label{sec:observations}
\begin{figure*}
  \centering
  \begin{minipage}[b]{0.49\textwidth}
    \includegraphics[width=\textwidth]{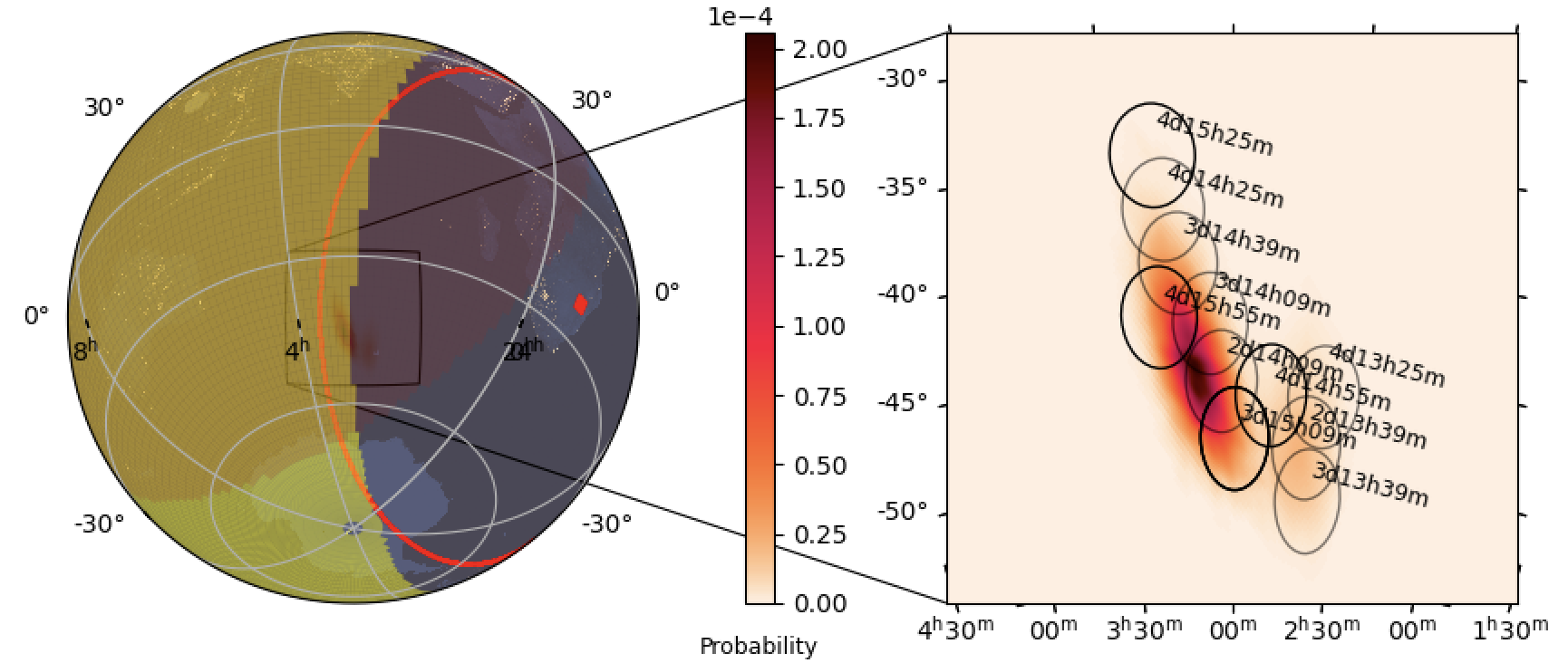}
  \end{minipage}
    \begin{minipage}[b]{0.49\textwidth}
    \includegraphics[width=\textwidth]{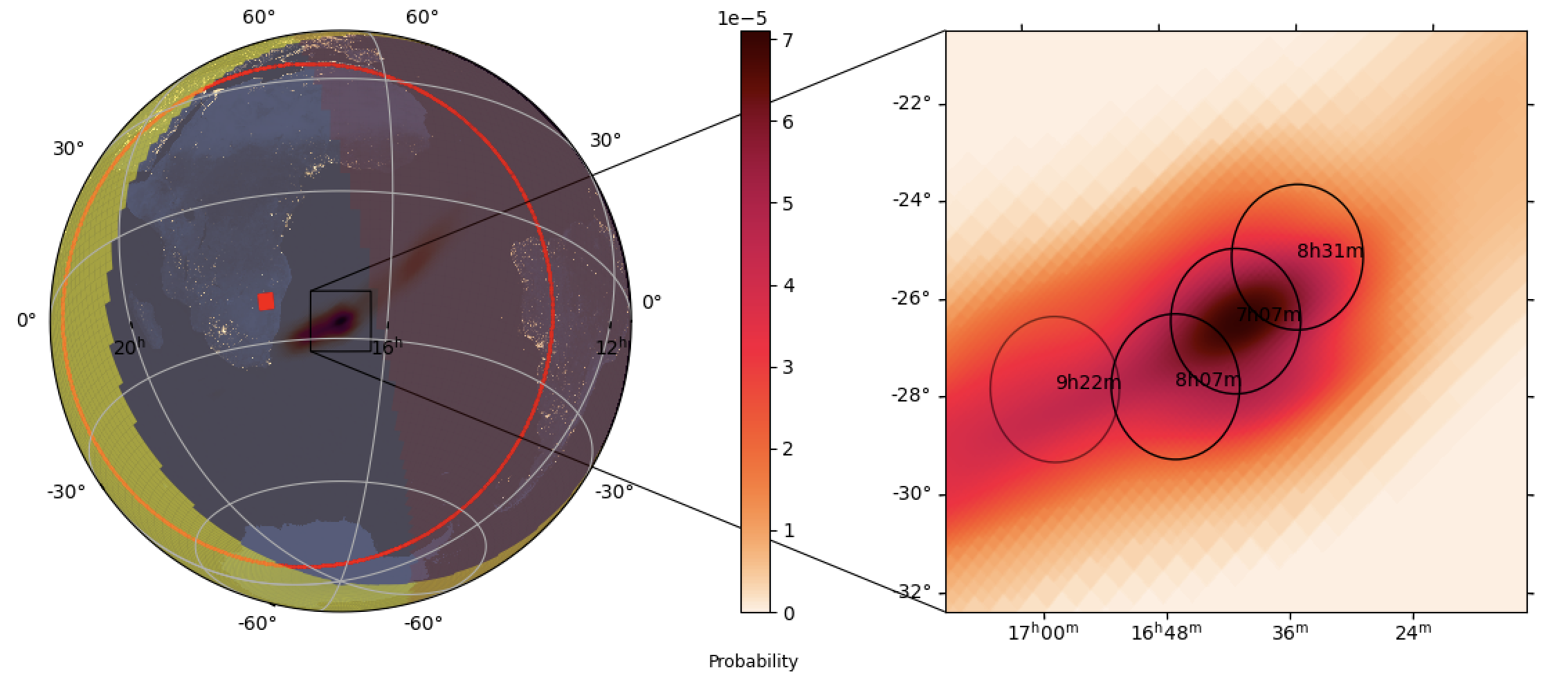}
  \end{minipage}
    \begin{minipage}[b]{0.49\textwidth}
    \includegraphics[width=\textwidth]{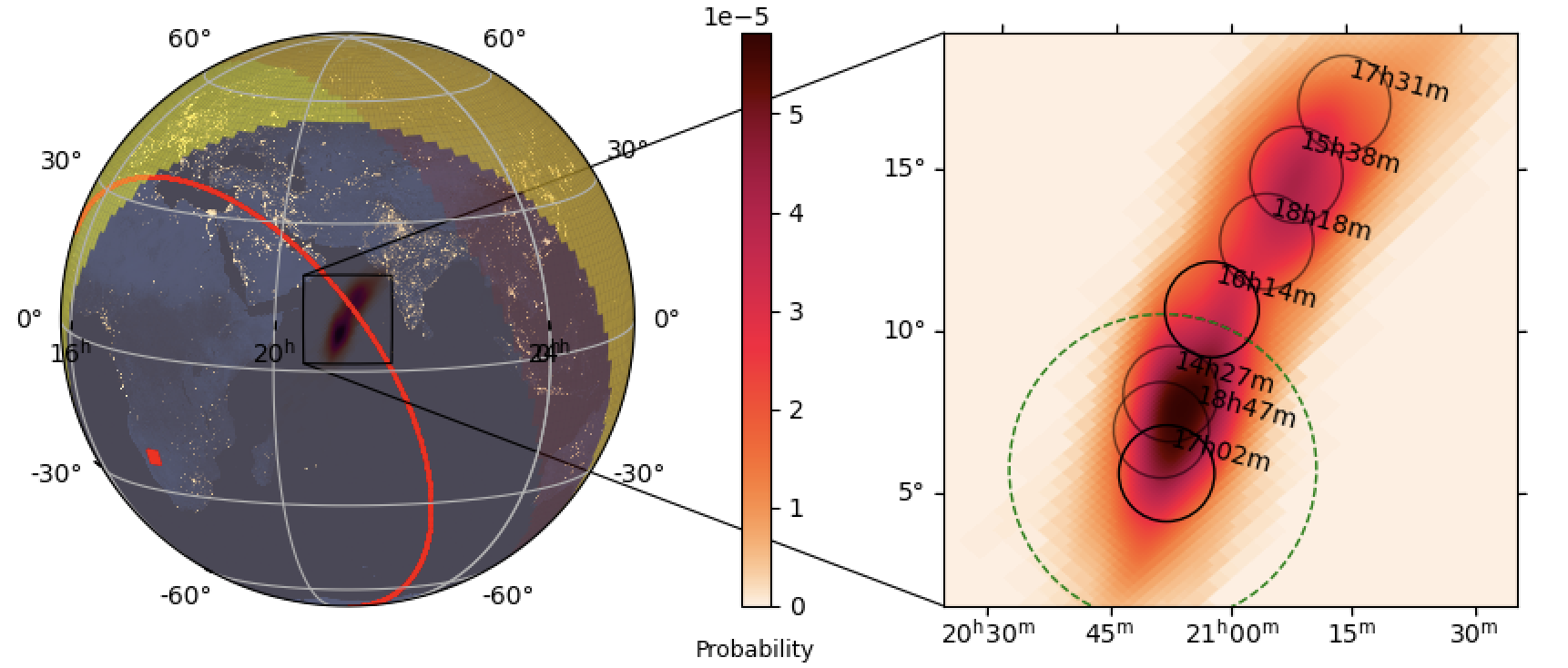}
  \end{minipage}
     \begin{minipage}[b]{0.49\textwidth}
    \includegraphics[width=\textwidth]{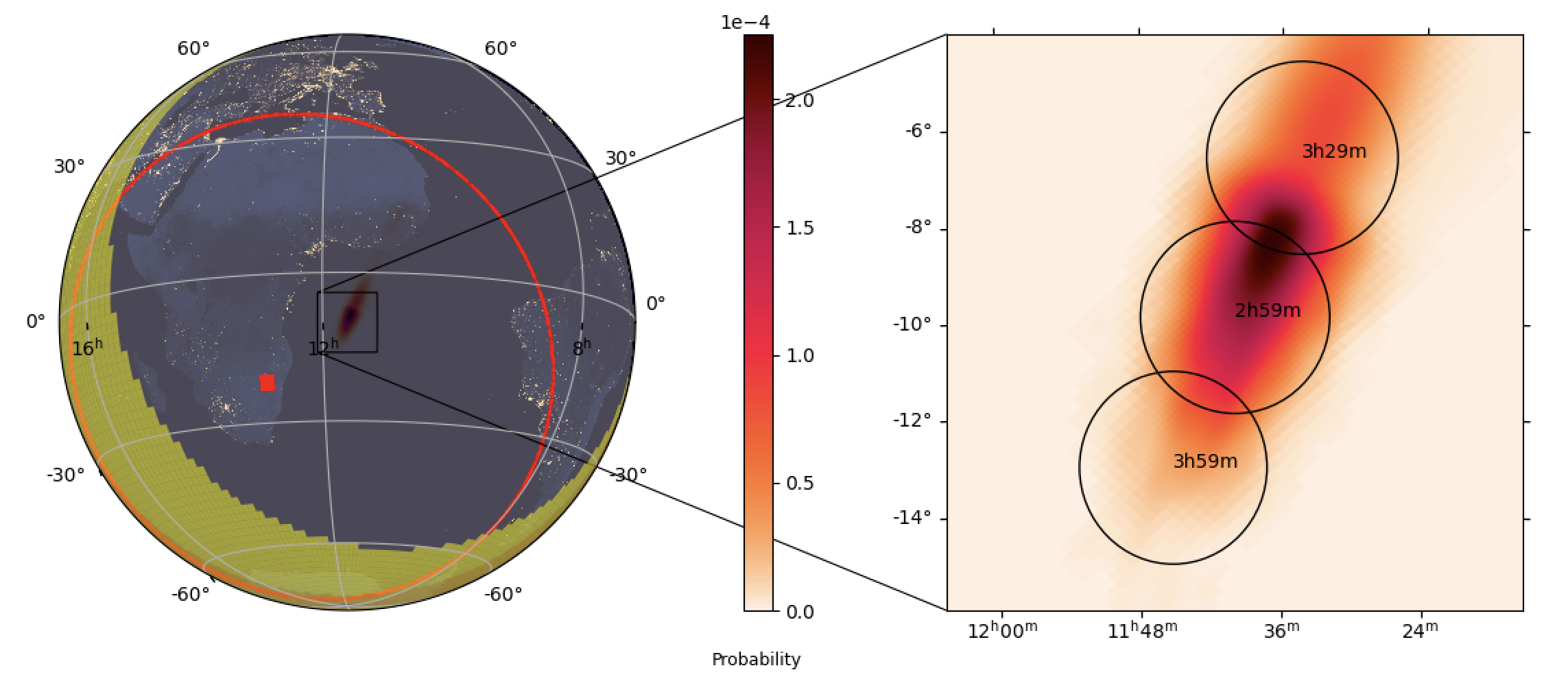}
  \end{minipage}
      \caption{H.E.S.S. coverage of four BBH events during O2 and O3: GW170814 (upper left), S190512at (upper right), S190728q (lower left), S200224ca (lower right). \textbf{Astro glob view}: The colors of the colorbar indicate the localisation probability. The Earth is seen in the background from the GW event point of view at the time of the beginning of observations. The yellow and brown patches indicate the regions on Earth where the sun's altitude is $>$ -18 and the moon's altitude is $>$ -0.5 deg respectively. The red square indicates H.E.S.S. location on Earth and the red line indicates the limit where the zenith angle is $<$ 60 deg. \textbf{Inset view}: The H.E.S.S. field of view of the observations taken on the GW events are presented by the black circles. The dark circles indicate observations taken at zenith angles $<$ 40 deg while the liter circles indicate observations taken at zenith angles $>$ 40. The field of view radius is 1.5 deg for S190512at and S190728q, 2 deg for S200224ca and 2.5 deg for GW170814. The observation delay for each pointing is shown in days, hours and minutes. For S190728q the green dashed circle represent the neutrino uncertainty region.}
    \label{fig:HESS_OBS_GW}
\end{figure*}

The large localisation uncertainties of GW events require specific observation strategies that relies on \textit{tilling} the localisation region in order to maximize the probability of observing the event. In fact, localisation maps are provided by the GW interferometers that contain the probability and distance information of finding the GW event in a region in the sky. Each region with a high probability of hosting the event is observed with 1 observation run. If a signal is detected in the real-time analysis, H.E.S.S. can choose to spend more time on the region. The follow-up strategies are developed for H.E.S.S. and are extensively described in~\cite{Technical_paper} and~\cite{technical_proceeding} alongside the pointing pattern of the observations described here. The requirement for BBH merger observations is that at least 50\% of the localisation map is covered by the observations. 
The observations are illustrated in Fig.~\ref{fig:HESS_OBS_GW}. The H.E.S.S. scheduling was obtained with the initial GW localisation maps distributed by notices, while here, the final published ones are shown. 

\section{Data analysis}
\label{sec:data_analysis}
For each follow-up observation, the data undergoes a quality check described in~\cite{Crab_HESS}. Only the data from the four 12-m telescope is used since a new camera for the 28-m telescope was installed and was under commissioning around the end of 2019 and beginning of 2020. The selection of the gamma-candidate events is performed with the \textit{Model} analysis described in~\cite{denorois_model}. The background subtraction is performed using the "ring background" technique from~\cite{Berge_RingBack} where the background is determined from the data itself and the acceptance of the events in the camera is assumed to be radially symmetric. Maps are created with a 0.02 deg pixel size. The background subtraction technique is performed on each bin of the map with a circular ON region of 0.1 deg radius. Only the bins with statistics $\mathrm{\alpha\, N_{OFF} > 5}$ are used (see~\cite{Berge_RingBack}). Excess maps are created and the excess is transformed to significance using the formalism described in~\cite{LIMA}.  No significant signal is found in the data. This result is verified with a separate independent analysis using the \textit{Image Pixel-wise fit for Atmospheric Cherenkov Telescopes} (ImPACT)~\cite{Parsons2014} software. 

Integral upper limit maps are derived following~\cite{Galactic_plane_survey}. An $\mathrm{E^{-2}}$ spectrum is assumed and integral upper limits are derived between 1 and 10 TeV which corresponds to the H.E.S.S. core energy range. These integral upper limit maps are used to derive luminosity upper limits of VHE gamma rays from the source. The distance estimation of a GW event can vary by several hundreds of Mpc in the region observed by H.E.S.S. To account for that, the per-pixel distance is considered to derive the luminosity upper limits.

To constrain emission on Earth, the absorption by the extragalagtic background light (EBL) should be taken into consideration. For these measurements, the maximum energy range reachable by the telescopes is considered. The minimum and maximum energy is considered where energy reconstruction biases are less than 10\%. For the minimum energy, $\mathrm{E_{th}}$, the value is increased to values where the acceptance is at $10\%$ of its maximum value. For the spectrum shape, the equivalent power law spectral index at the $\mathrm{E_{th}}$ of each event is considered by assuming an EBL absorption model~\cite{Franceschini} at the redshift of the GW event. The usage of different EBL absorption models and different energies  than $\mathrm{E_{th}}$ up to 1 TeV, translates into a difference of less than 10\% in the upper limits derived. The analysis is detailed in~\cite{HESS_BBH_RESULTS} and the spectral indices and energy ranges are given in Tab.~\ref{tab:GW_REDSHIFT_INDEX_COV} alongside the achieved effective coverage for each GW event.  
\begin{table*}[ht!]
    \centering
    \small
    \begin{tabular}{ccccc}
     \hline
    GW event &  Redshift & $\gamma(E = E_{th}$, $z= z_{GW})$ & E (TeV) & Coverage\\ 
    \hline
    GW170814 & 0.12 & 2.73 & 0.42-34.80 & 75.4\%\\
    S190512at & 0.28 & 3.58 & 0.31-38.31 & 34.5\%\\
    S190728q & 0.18 & 2.98 & 0.35-26.10 & 50.8\%\\
    S200224ca& 0.29 & 3.08 & 0.24-38.31 & 62.13\%\\
    \hline
    \end{tabular}
    \caption{Spectral indices, $\gamma$ (col 3) at the GW event corresponding redshift (col 2) and at $E_{th}$ assuming a $E^{-2}$ source, the energy range used to derive the specific integral upper limit maps (col 4) and the corresponding probability coverage (col 5). From~\cite{HESS_BBH_RESULTS}.}
    \label{tab:GW_REDSHIFT_INDEX_COV}
\end{table*}
The specific integral upper limit maps are used to derive upper limits on the observed energy flux.  The map-averaged energy flux and luminosity values are presented in Tab.~\ref{tab:flux_lum_ULs}.

\begin{table*}[ht!]
    \centering
    \small
    \begin{tabular}{ccccccc}
    \hline
    GW event & & \multicolumn{2}{c}{energy flux, event-specific (erg cm$^{-2}$ s$^{-1}$)} & & \multicolumn{2}{c}{luminosity, standard (erg s$^{-1}$)}\\\cline{3-4}\cline{6-7}
    & & mean & standard dev & & mean & standard dev \\
    \hline
    GW170814 && $3.7\times10^{-12}$ & $1.8\times10^{-12}$ & & $1.3\times10^{44}$ & $9.8\times10^{43}$\\
    S190512at && $3.1\times10^{-12}$ & $1.5\times10^{-12}$ & &$9.9\times10^{44}$ & $4.7\times10^{44}$\\
    S190728q && $2.6\times10^{-12}$ & $1.3\times10^{-12}$ & &$3.2\times10^{44}$ & $1.6\times10^{44}$\\
    S200224ca&& $2.7\times10^{-12}$ & $1.2\times10^{-12}$ & &$1.9\times10^{45}$ & $8.8\times10^{44}$\\
    \hline
    \end{tabular}
    \caption{The energy flux and luminosity upper limits for the four GW events calculated individually for each pixel in the sky region observed by H.E.S.S. then averaged. The energy flux upper limits here are EBL absorbed and are calculated over the event-specific energy ranges in Tab.~\ref{tab:GW_REDSHIFT_INDEX_COV}. The luminosity are calculated from the unabsorbed energy fluxes assuming an $E^{-2}$ source spectrum and a 1--10 TeV energy range, and using the per-pixel luminosity distances. These values are also plotted in Figs.~\ref{fig:luminosity_comparisons}. From~\cite{HESS_BBH_RESULTS}}
    \label{tab:flux_lum_ULs}
\end{table*}

\section{Discussion}
\label{sec:discussion}
A coverage of the localisation uncertainty greater than 50\% is achieved with the H.E.S.S. pointing pattern except for S190512at. Therefore the derived upper limits can be assumed as spatially constraining. The question now is, how much the upper limits of the H.E.S.S. instruments using the tiling pointing pattern (described in Sec.~\ref{sec:observations}) that maximizes coverage are constraining. To assess this, the upper limits derived here are compared to the VHE emission of H.E.S.S detected GRBs and to extrapolated emission from \textit{Fermi}-LAT~\cite{LAT_GRBs} GRBs with known redshifts and extended emission. 

\begin{figure*}[ht!]
  \centering
    \includegraphics[width=0.49\textwidth]{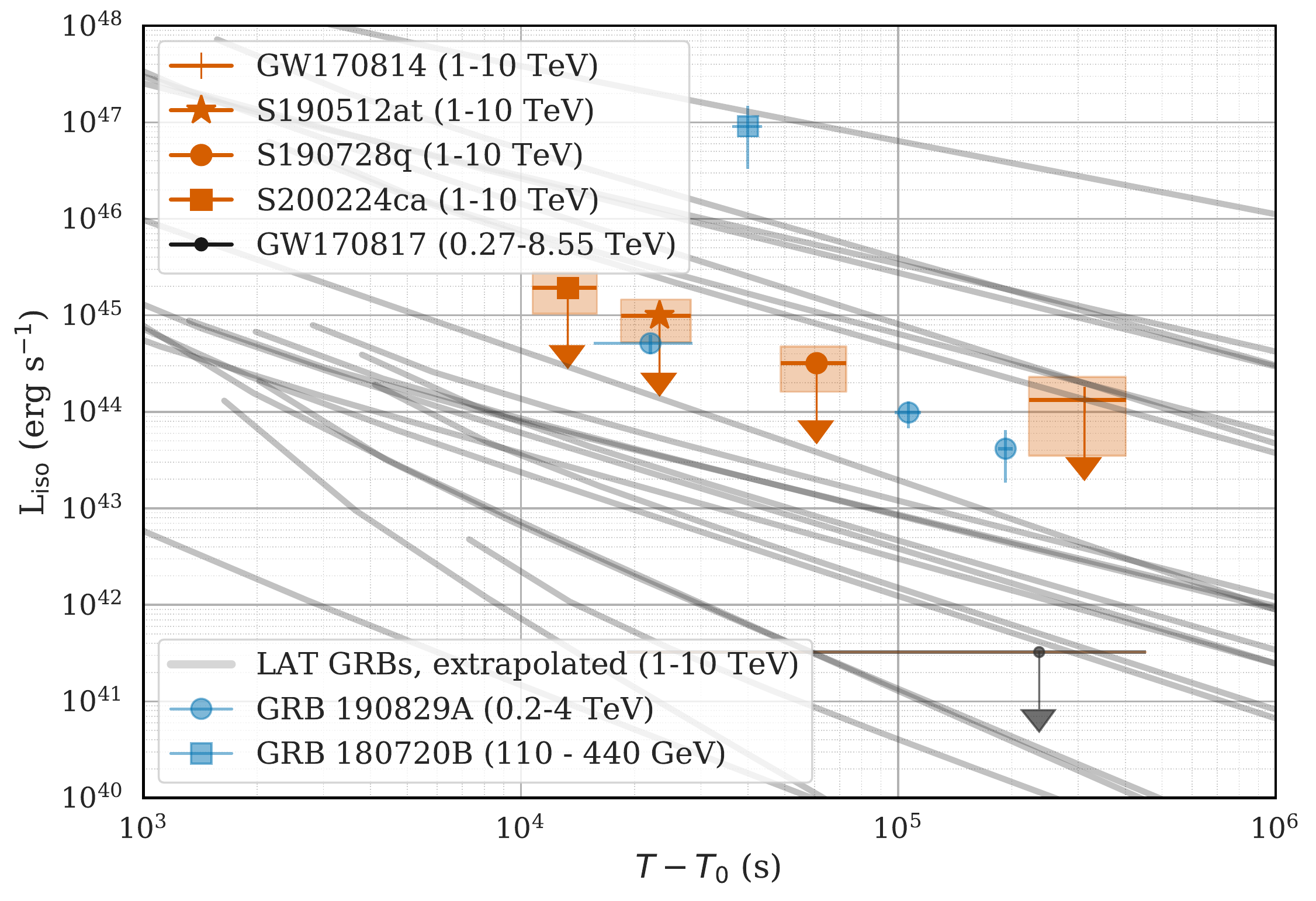}
    \includegraphics[width=0.49\textwidth]{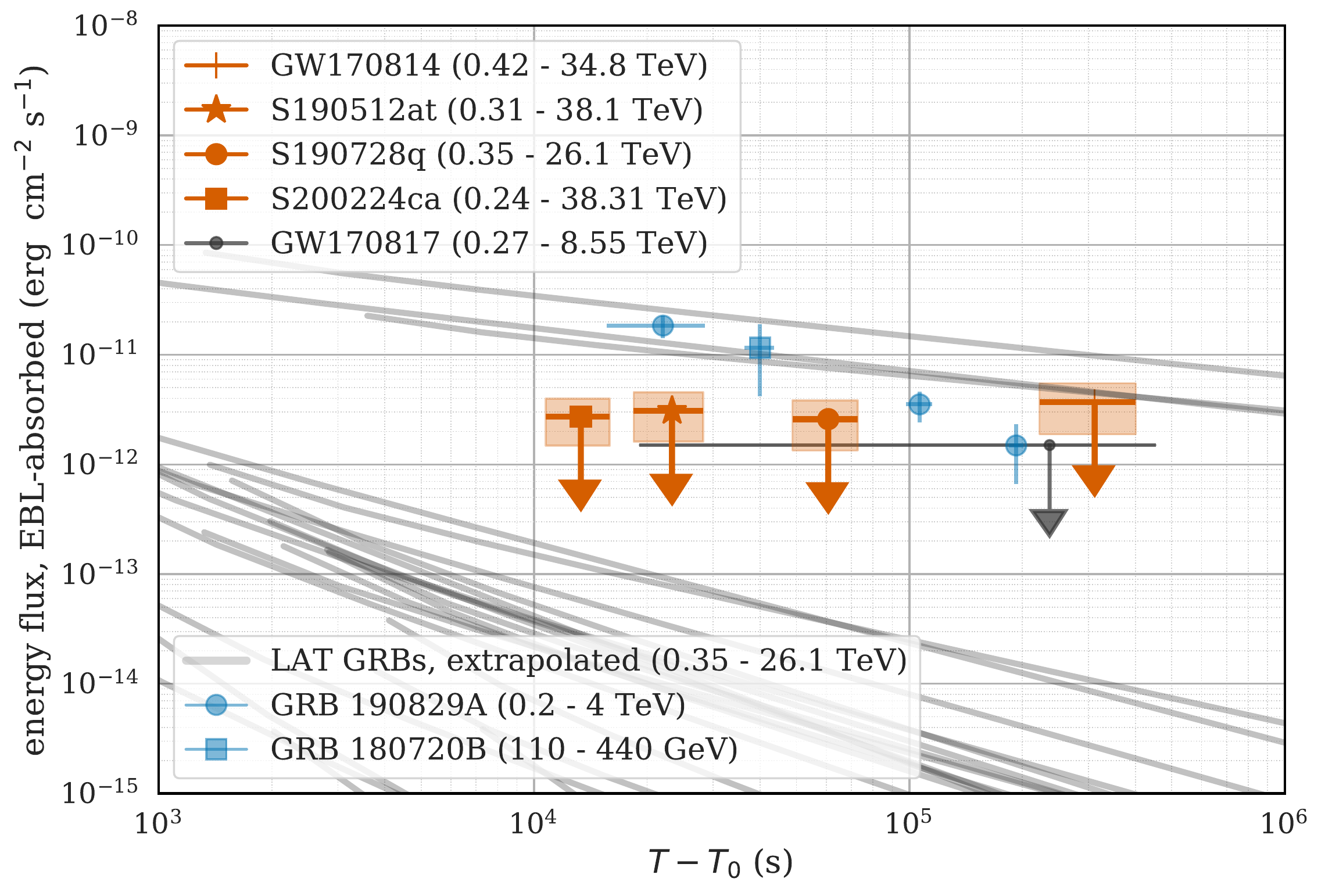}
    \caption{Mean (orange points) and standard deviation (orange bands) of the per-pixel luminosity upper limit maps (right) and and the integral energy flux upper limits (left) for the BBH events. The luminosity upper limits are of all five GW upper limits are calculated assuming an intrinsic $E^{-2}$ spectrum, although the upper limit for GW170817 is calculated with a slightly different energy range. The integral energy flux upper limits are calculated from the per-pixel EBL-absorbed integrated energy flux upper limit maps. These are calculated using the specific energy range and index for each event, based on the redshift in Tab.~\ref{tab:GW_REDSHIFT_INDEX_COV}.  The luminosity upper limits are compared to luminosity extrapolation of Fermi-LAT GRBs (grey lines) with known redshift, The luminosity from H.E.S.S. detected VHE GRBs and to the H.E.S.S. upper limit on GW170817 (black)~\cite{GW170817_HESS} and the intergral energy flux upper limits are compared to the energy flux extrapolations of LAT GRBs (grey lines) into the specific energy band of S190728q and, to the EBL-absorbed energy flux of H.E.S.S. detected VHE GRBs and to the GW170817 energy flux upper limits. The EBL absorption for the GRBs is calculated based on the redshift of S190728q. From~\cite{HESS_BBH_RESULTS}.}
\label{fig:luminosity_comparisons}
\end{figure*}

We start by comparing the luminosity upper limits. For the \textit{Fermi}-LAT GRBs, the spectrum measured by the LAT at late times in the 100 MeV to 100 GeV band is extrapolated into the H.E.S.S. energy bands (1-10 TeV) using the spectral index measured by the LAT at these times. The emission is then extended in time using the power-law decay index measured by the LAT at late-time. The energy flux is then converted to isotropic luminosity at the GRB redshifts. The detected H.E.S.S. VHE GRBs, GRB\,180720B~\cite{GRB180720B_HESS} and GRB\,1908929A~\cite{GRB190829A_HESS} are also shown for comparison. Their detection time is similar to the observation delays of the GW events observed with H.E.S.S. Their EBL corrected energy flux extrapolation is also converted into isotropic luminosity. The isotropic luminosity from the LAT extrapolations and the VHE GRBs are compared in Fig.~\ref{fig:luminosity_comparisons} to the H.E.S.S. upper limits. We also include the upper limits derived from the GW170817 short term observations~\cite{GW170817_HESS}.  
From the left plot of Fig.~\ref{fig:luminosity_comparisons}, we see that the upper limits derived from the H.E.S.S. observations lie below some of the extrapolated LAT GRB level. They are also at the same level of GRB\,190829A which is at a similarly low redshift and below the GRB\,180720B level which is at a relatively high redshift. The level of the upper limits derived from the GW170817 H.E.S.S. observations is three orders of magnitude lower than the upper limits presented in this study. This is due to the proximity of this event. This suggests that if the studied GW events produced GRBs similar to the ones shown in the plot, H.E.S.S. would have a good chance of detecting the VHE emission. 

To compare the observed energy flux, we use the extrapolation of the LAT GRBs mentioned earlier taking into account how the EBL absorption will affect the extrapolated emission. The assumed redshift is the one for S190728q; choosing the redshift of one of the other four BBH GW events decreases the energy flux extrapolations by less than 50\% for S190512at and S200224ca, and increases by less than 75\% for GW170814. For the two H.E.S.S. detected VHE GRBs: We calculate the energy flux of the observed spectrum using the power law fit to the EBL-attenuated data for GRB\,180720B. For GRB 190829A, we calculate the EBL-attenuated energy flux for each of the three nights separately using the constant intrinsic photon index (2.07) derived by combining the data from all three nights since this information is not available for the third night alone. 

The results are shown in the right plot of  Fig.~\ref{fig:luminosity_comparisons}. The upper limits for the four BBH merger events studied here span a large range in observation delays, but lie at similar levels. We can see that in this case, the H.E.S.S observations would have to have happened at a much earlier phase (with smaller $T- T_0$) or spend more time on a single sky position to increase the sensitivity. For the latter, longer observations per target would be required. Unlike for GRBs, during GW follow-up observations no single sky position gets as much exposure as it would in a standard single-position multi-hour follow up. However, looking at the levels of the GW170817 upper limits, which were derived with 3.2 hours of observation, we can see that they lie below the level of the H.E.S.S. sensitivity presented here but are still not at the level of the extrapolated LAT GRBs. Therefore, focusing on minimizing $T- T_0$ would lead to better results.

Finally, the assumptions taken here for the LAT GRBs is that the spectral shape and the temporal evolution remains unchanged. Therefore, the LAT extrapolations should be considered as simply representing a \emph{range} of potential behavior and should not be considered as precise predictions. Moreover, the temporal extrapolation assumes that all the LAT GRBs are on-axis events. This assumption holds given that the light curves of these GRBs decays with time and no subsequent re-brightening was observed like in the case of GW170817.

\section{Prospects for O4 and conclusion}
\label{sec:prospects}
All the observation delays achieved with the H.E.S.S. observations of GW events during O2 and O3 are on the order of hours (days for GW170814). This is due to the necessity of waiting for favourable observation conditions (darkness, low zenith angle, good weather and cloudless sky) to occur. Given the rate of GW events detected in O3 and the large localisation uncertainties, H.E.S.S. was unlikely to have observed an event with good localisation with favourable zenith angle and time delay. In fact, for O3, the expected number of events with a good localisation (similar to the events studied here) was around 25\%. For an expected event rate of $18^{+53}_{-12}$ per year~\cite{LVKC_prospects}, this number becomes $4^{+13}_{-3}$. Considering that only half the sky is reachable by H.E.S.S. and assuming an isotropic distribution of the GW events in the sky, this number drops to $2^{+7}_{-1}$ events per year that would qualify for H.E.S.S. follow-up. Given that H.E.S.S. can observe on average 6 hours per night (averaging over an entire year) only $0.5^{+1.5}_{-0.5}$ events would have been observable during H.E.S.S. observation time, which would have allowed an automatic prompt follow-up. This expectation turns out to be accurate since H.E.S.S. could only observe S20022ca promptly during O3 (which did not happen due to weather conditions).

However, the situation is expected to change for O4. With a prediction of $90^{+232}_{-55}$ GW events per year, of which 75\% are expected to have small localisation uncertainties that will pass the H.E.S.S. follow-up criteria, we would have $67^{+174}_{-41}$ mergers per year. Assuming an isotropic distribution in the sky and that around half the sky is reachable  by H.E.S.S. again, this number is divided by 2. Finally, assuming an average of 6 hours of observation per night, we expect that $8^{+22}_{-5}$ events per year will occur during H.E.S.S. observation time (with prompt follow-ups). These prompt follow-ups are guaranteed to reduce $T - T_0$ from Fig.~\ref{fig:luminosity_comparisons} to less than 10 minutes (assuming a few minutes latency for the distribution of the GW alert). Moreover, 35\% of the O4 events are expected to have a 50\% localisation uncertainty smaller than few deg$^2$ in the sky, which means that they could be covered by H.E.S.S. with at least one pointing. This means that H.E.S.S. can spend more time on deeper observations for 1 position instead of scanning the localisation region with 1 pointing per position. This would then allow the energy flux upper limit in Fig.~\ref{fig:luminosity_comparisons} to improve by a factor of $\sim$\,4 depending on the amount of observation time and assuming a maximum continuous observation time of 8 hours. The number of events qualifying for this type of observations is $8^{+20}_{-5}$ per year. Finally, the number of events that would qualify for prompt observations and can be observed with only one pointing is $4^{+10}_{-2}$ per year.

In conclusion, we find that minimizing the observation delay (i.e., following up GW events that are immediately observable by H.E.S.S.) would have a greater effect on the detectability than reducing the sky coverage and spending more time observing single positions. However, achieving both will lead to better results. This will naturally happen in the next observing runs, with the increased rate of GW detections. Therefore, we do not expect to fundamentally alter our observing strategy, and will continue to prioritize sky coverage. 

\section*{Acknowledgements}
The H.E.S.S. acknowledgements can be found in: \url{https://www.mpi-hd.mpg.de/hfm/HESS/pages/publications/auxiliary/HESS-Acknowledgements-2021.html}


\clearpage
\section*{Full Authors List: \Coll\ Collaboration}


\scriptsize
\noindent
H.~Abdalla$^{1}$, 
F.~Aharonian$^{2,3,4}$, 
F.~Ait~Benkhali$^{3}$, 
E.O.~Ang\"uner$^{5}$, 
C.~Arcaro$^{6}$, 
C.~Armand$^{7}$, 
T.~Armstrong$^{8}$, 
H.~Ashkar$^{9}$, 
M.~Backes$^{1,6}$, 
V.~Baghmanyan$^{10}$
% ...

\medskip

\noindent
$^{1}$University of Namibia, Department of Physics, Private Bag 13301, Windhoek 10005, Namibia\\
$^{2}$Dublin Institute for Advanced Studies, 31 Fitzwilliam Place, Dublin 2, Ireland\\
$^{3}$Max-Planck-Institut f\"ur Kernphysik, P.O. Box 103980, D 69029 Heidelberg, Germany\\
$^{4}$High Energy Astrophysics Laboratory, RAU,  123 Hovsep Emin St  Yerevan 0051, Armenia\\
$^{5}$Aix Marseille Universit\'e, CNRS/IN2P3, CPPM, Marseille, France\\
$^{6}$Centre for Space Research, North-West University, Potchefstroom 2520, South Africa\\
$^{7}$Laboratoire d'Annecy de Physique des Particules, Univ. Grenoble Alpes, Univ. Savoie Mont Blanc, CNRS, LAPP, 74000 Annecy, France\\
$^{8}$University of Oxford, Department of Physics, Denys Wilkinson Building, Keble Road, Oxford OX1 3RH, UK\\
$^{9}$IRFU, CEA, Universit\'e Paris-Saclay, F-91191 Gif-sur-Yvette, France\\
$^{10}$Instytut Fizyki J\c{a}drowej PAN, ul. Radzikowskiego 152, 31-342 Krak{\'o}w, Poland\\
% ...

\end{document}